# Improvement of FTJ on-current by work function engineering for massive parallel neuromorphic computing

Suzanne Lancaster
*NaMLab gGmbH,*
Dresden, Germany
suzanne.lancaster@namlab.com

Quang T. Duong
*NaMLab gGmbH,*
Dresden, Germany
quang.duong@namlab.com

Erika Covi
*NaMLab gGmbH*
Dresden, Germany,
erika.covi@namlab.com

Thomas Mikolajick
*NaMLab gGmbH & IHM TUD*
Dresden, Germany
thomas.mikolajick@namlab.com

Stefan Slesazeck
*NaMLab gGmbH*
Dresden, Germany
stefan.slesazeck@namlab.com

*Abstract*— $HfO_2$-based ferroelectric tunnel junctions (FTJs) exhibit attractive properties for adoption in neuromorphic applications. The combination of ultra-low-power multi-level switching capability together with the low on-current density suggests the application in circuits for massive parallel computation. In this work, we discuss one example circuit of a differential synaptic cell featuring multiple parallel connected FTJ devices. Moreover, from the circuit requirements we deduce that the absolute difference in currents $I_{on} - I_{off}$ is a more critical figure of merit than the tunneling electroresistance ratio (TER). Based on this, we discuss the potential of FTJ device optimization by means of electrode work function engineering in bilayer $HZO/Al_2O_3$ FTJs.

## I. INTRODUCTION

In non-von-Neumann architectures the latency and power consumption required for computational tasks is reduced by repealing the separation between memory and computing engines. Instead, data is processed where it is stored [1]. Especially in power-efficient edge computing, the ability for normally-off computing requires the utilization of non-volatile memory devices, thus enabling the system to be powered on and off frequently to process data only when needed or to cope with frequent power interruptions. In recent years the ferroelectric tunnel junction (FTJ) has emerged as an interesting non-volatile memory device featuring non-destructive readout [2], high-speed write operation [3], low voltage operation, and high endurance cycling [4]. The concept of an FTJ was first introduced by Esaki et al. [5]. The operating principle is based on a difference between potential barrier height and width at a metal-ferroelectric interface depending on the direction of the spontaneous polarization (P). This polarization can be switched using an external electric field. According to Simmons [6], the tunneling current density is inversely proportional to the potential barrier, with

$$J \sim \exp(-\sqrt{\varphi}), \qquad (1)$$

where $\varphi$ is the potential barrier height above the Fermi level. Therefore, two distinct resistance states could be generated depending on the direction of the polarization. The difference in conductance of the two states is called the tunneling electroresistance ratio (TER), given by TER=$(G_{on}-G_{off})/G_{off}$, where $G_{off}$ is the conductance for the $P_{down}$ state and $G_{on}$ is the current for the $P_{up}$ state [7].

Recently, a differential synaptic cell featuring two FTJ devices has been proposed [8] as is depicted in Fig. 1 a. The circuit concept is based on a combination of a current normalizing differential synaptic weighting element [9] and a 2T1C read-out circuit to cope with the small on-currents of small-scaled FTJ devices [10]. In this circuit the readout is performed in two steps by first pre-charging the gates of $T_3$ and $T_4$ via the access transistors to bit-line (BL) while already applying the readout voltage difference between plate-line (PL) and BL, and in a second phase the access transistors are switched off in order to develop a voltage difference $\Delta V$ between nodes $n_1$ and $n_2$ by charging the gates of $T_3$ and $T_4$ via the FTJ devices $C_1$ and $C_2$.

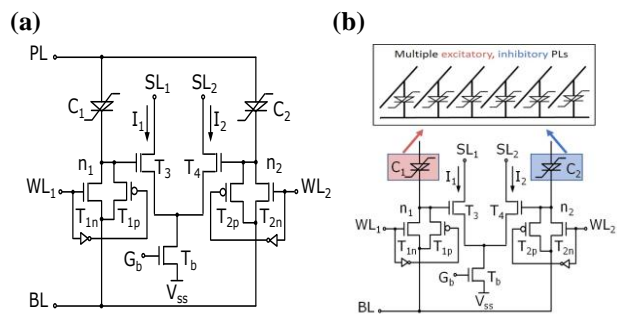

Fig. 1. (a) Differential synaptic cell using two FTJ devices. © 2021 IEEE. Reprinted, with permission, from [8]; (b) differential synaptic cell using multiple parallel FTJ devices.

As has been discussed already in [8] and [11], the voltage difference $\Delta V$ depends mainly on the FTJ current density $J_{FTJ}$, the FTJ self-capacitance per area $C_{0,FTJ}$ and the signal development time $t_{read}$ according to the equation:

$$\Delta V = (J_{FTJ}\ t_{read})/C_{0,FTJ} \qquad (2)$$

Moreover, it has been shown by means of circuit simulation using calibrated FTJ models, that despite the very low on-current densities of the used FTJ devices in the range of <1 pA/µm² this synaptic weighting cell can be operated successfully at biologically plausible time scales with read times in the 100 µs range and typical values of $\Delta V$ in the range of 100 mV. In this simplest version of the synaptic cell 7 transistors are used. The voltage signals to operate the cell such as PL, BL, word-line (WL), and its complement voltages are provided from outside and the respective transistors are not shown. Hence, when looking from a system level perspective on this circuit concept it turns out that the number of transistors per synaptic weighting element is large, which makes the design not very efficient, both in terms of silicon area as well as power consumption.

## II. PARALLEL FTJs IN DIFFERENTIAL SYNAPTIC CELL

Fig. 1 b depicts a differential synaptic cell that uses multiple parallel connected FTJs. The idea of this concept is to integrate multiple FTJ devices in a cross-bar array on top of the CMOS circuitry. Thus, multiple FTJ devices could be read in parallel, thereby greatly reducing the CMOS circuitry overhead. This configuration also allows the implementation of multi-level cells by parallel operation of multiple single-level cells that can eventually reduce the programming circuit overhead. Another interesting aspect arising for a large number, n, of FTJs is that the overall node capacitance connected to $n_1$ and $n_2$ increases by a factor of n with respect to the capacitance of the individual FTJ. That enables to perform program and erase operations of single FTJs without the need to connect the nodes $n_1$ and $n_2$ actively to the BL, but rather simply using the capacitive voltage divider between one active and (n - 1) passive FTJs by activating the respective PLs. Hence, the access transistor would be used for pre-charge during read operation only, which allows to get rid of the pFET access transistors ($T_{1p}$, $T_{2p}$) and removes the need for generation of complementary WL signals from the circuit in Fig. 1a.

Due to the increased capacitance of the nodes $n_1$ and $n_2$ by a factor n, the readout of single FTJs in a kind of single-bit access will not be possible anymore since according to (2) the $\Delta V$ scales with 1/n. However, the case is different for the parallel readout as one would envisage for typical non-von Neumann architectures: the dot product can be realized by applying voltage pulses to individual PLs, the pulse time representing a first analogue value, and weighting this input signal by the conductivity of the individual FTJs, finally accumulating the resulting charge on $n_1$ and $n_2$. Another implementation could be using just digital representation of data in large vectors, which would be interesting for hyper dimensional computing (HDC) concepts [12].

Interestingly, in this parallel configuration of multiple FTJs the maximum voltage difference $\Delta V$ that develops during read phase at the gates of $T_3$ and $T_4$ would not be altered compared to the case of a single FTJ, since both current density as well as capacitance would scale linearly with the number of FTJs:

$$\Delta V = (J_{FTJ}\ t_{read})\ /\ C_{0,FTJ} = (n\ J_{FTJ}\ t_{read})\ /\ (n\ C_{0,FTJ}). \quad (3)$$

Hence, after readout $\Delta V$ represents the weighted sum of all inputs, and in addition the natural limit of $\Delta V$ that is given by (3) realizes a window function that, depending on the interpretation of the resulting difference in currents $I_1$ and $I_2$ could be further interpreted as a linear approximation of a tanh or sigmoid activation function.

In summary, the proposed differential cell featuring multiple parallel FTJs for realization of multiple weighting elements offers many advantages. However, in order to make the readout less susceptible to noise, for a given FTJ technology $\Delta V$ can be only increased in the trade-off against increasing the readout time $t_{read}$. Hence, for device engineering one of the most important targets is still to increase the attainable current density $J_{FTJ}$ while keeping the capacitance per area $C0_{FTJ}$ low. In addition, the above argumentation highlights that in our readout concept the absolute current difference between $I_{ON}$ and $I_{OFF}$ is a much more important figure of merit than the oft-discussed TER.

## III. ELECTRODE WORK FUNCTION ENGINEERING

### A. Sample preparation

According to (1) the current density of the FTJ can be increased by lowering the effective potential barrier at the injecting electrode. In the case of our double layer FTJ [7] injection takes place from the top electrode which is in contact with the $Al_2O_3$ layer. This can be attained by the reduction of the work-function by replacing the TiN TE (work function of 4.7 – 5.2 eV [13]) by either Al (work function of 4.0 – 4.2 eV) or by doping Al into the TiN TE.

In our experiments, three different stacks have been prepared for direct comparison: TiN/$Hf_{0.5}Zr_{0.5}O_2$/$Al_2O_3$/TiN (referred to as *standard stack*), TiN/$Hf_{0.5}Zr_{0.5}O_2$/$Al_2O_3$/Al (referred to as *Al top electrode stack*), and TiN/$Hf_{0.5}Zr_{0.5}O_2$/$Al_2O_3$/[TiN/TiAl/TiN] (referred to as *Al-doped TiN top electrode stack*).

At first, starting with a mid- doped Si wafer, 30 nm of tungsten (W) is deposited by physical vapor deposition (PVD) at room temperature. Next, a 10 nm TiN bottom electrode is deposited at room temperature by sputtering under high vacuum conditions. The ferroelectric film $Hf_{0.5}Zr_{0.5}O_2$ (HZO) with a nominal thickness of 10 nm is grown by atomic layer deposition (ALD) at 280 °C with a 1:1 ratio of $HfO_2$ and $ZrO_2$ with ozone as oxidant. Similarly, a 2 nm $Al_2O_3$ layer is also deposited by ALD at 280 °C, using TMA as precursor and ozone as the oxidant source.

From this point, three different top electrodes have been deposited to complete the structure. The first sample utilizes Al top electrode with a thickness of approximately 25 nm. After aluminium oxide deposition, and before TE deposition, the anneal process was performed, followed by the e-beam evaporation of Al using a shadow mask to form a final capacitor structure.

The second sample contains a TiN top electrode (TE) that has been deposited with a nominal thickness of 10 nm. In the last samples, the laminate structure TiN/TiAl/TiN is deposited by PVD, with three different nominal thicknesses of TiAl, that is 2.5, 5.0, and 7.5 nm in order to verify the effect of Al contents towards current density and ferroelectric properties. In both cases, the stacks undergo a crystallization anneal after TE deposition at 500 °C for 20 seconds in N2 atmosphere. A Ti/Pt dot was patterned using a shadow mask during e-beam evaporation, followed by the standard cleaning solution (SC-1, ammonium hydroxide and hydrogen peroxide) to etch the TiN electrode and define the capacitor area.

The final devices used for electrical characterization have diameters of 200 μm. The polarization-voltage
(P-V) hysteresis and cycling endurance were measured with a ferroelectric device tester (Aixacct TF3000). A Keithley

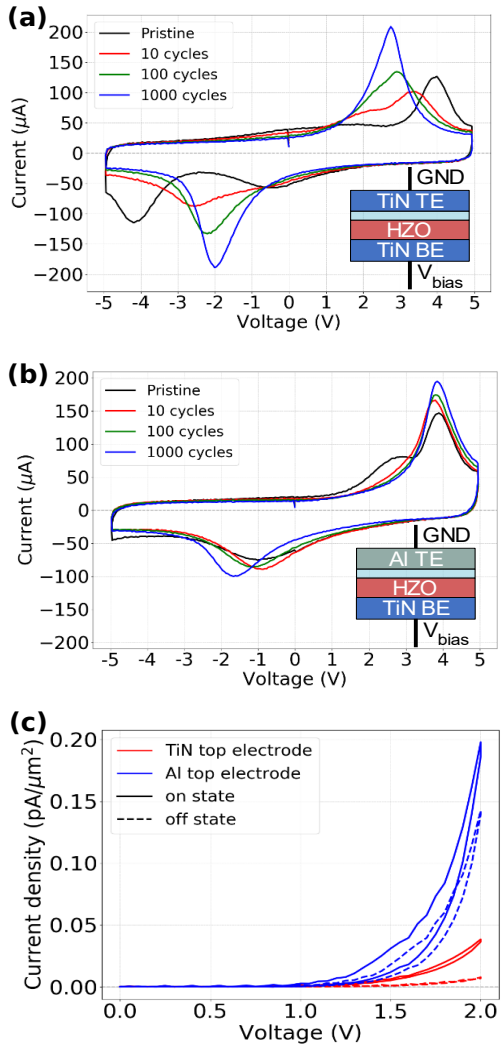

Fig.2. (a) I-V characteristics of TiN/Hf0.5Zr0.5O2/Al2O3/TiN sample (b) I-V characteristics of TiN/Hf0.5Zr0.5O2/Al2O3/Al measured at +/- 5 V, 1 kHz (cycling at 5 V, 10 kHz). Positive voltages indicate the $P_{up}$ state, and negative voltages indicate $P_{down}$. (c) Comparison of the current-voltage response of devices with different electrodes in the $P_{up}$ and $P_{down}$ state.

4200S was used to analyze the resistive switching behavior by obtaining the DC current-voltage response of the FTJ.

*B. Al top electrode*

In a first experiment the TiN TE was replaced by Al. Fig. 2 shows the comparison of the measured transient IV-switching characteristic and the DC-IV $I_{ON}/I_{OFF}$ currents of the FTJ devices. A wake-up effect can be observed clearly in the standard stack, merging two switching peaks into one peak [14]. In contrast, this effect is diminished in the Al top electrode stack, suggesting a more even distribution of oxygen vacancies directly after annealing [15]. Another feature is the shift of the hysteresis to the right in the Al top electrode sample, which means that a large voltage is needed to switch to the positive state ($P_{up}$), and the coercive voltage is smaller on the negative side ($P_{down}$), which can be explained due to the internal field caused by the work function difference between top and bottom electrodes [16]. This internal field may also be the cause for the redistribution of charges prior to electric field cycling.

In the standard process (TiN top electrode), after 1000 cycles, the switching voltages are -1.6 V and 2.3 V, with a remanent polarization $P_r = -22.61\ \mu C/cm^2$ and $+21.66\ \mu C/cm^2$, whereas, in the Al top electrode sample, these values are -1.5 V and 3.6 V, $P_r = -22.1\ \mu C/cm^2$ and $P_r = +15.94\ \mu C/cm^2$, respectively (as shown in Fig. 2). The overall larger coercive field of the Al-TE sample can be explained by an increase of the $Al_2O_3$ interlayer thickness by 0.9 nm due to partial oxidation of the Al-electrode during Al evaporation directly on $Al_2O_3$ and subsequent annealing, which was revealed by CV and XRR measurements (not shown). Despite the thicker tunneling barrier, when comparing with the DC-IV response of the standard stack we observed an increase in tunneling current by a factor of 4, with approximately 0.2 $pA/\mu m^2$ at 2 V readout for on-state, revealing the successful increase of the on-current by lowering the electrode work function. However, from reliability investigations we could identify that further oxidation of the top electrode is a critical degradation issue in the Al-TE based FTJ devices, strongly affecting the device's stability over time. Additionally, the strong built-in field impacts the stability of the two distinct resistance states and we calculated that due to fast polarization loss, only ~15% of switched domains contribute to the measured $I_{ON}$, meaning that the theoretical maximum $I_{OFF}$ could be much larger than shown here. In order to improve the stability of the devices, work-function engineered TEs were manufactured by doping Al into a TiN TE [17].

*C. Al-doped TiN top electrode*

The typical current-voltage (I-V) switching characteristic is shown in Fig. 3 a for three different TiAl insertion layers. Increasing the Al content in the top electrode increases the polarization $P_r$ value, whereas the coercive voltage $V_c$, is decreased. The increase in $P_r$ can be explained by a larger orthorhombic phase fraction as was revealed by measuring the grazing-incidence X-ray diffraction spectra of the HZO films (not shown). Furthermore, it is evidence that the work function of the TiN decreases with a thicker insertion TiAl layer, resulting in a higher work function difference between the bottom and top electrode. Therefore, a higher effective electric field is applied on the HZO film (in this case, the thickness of HZO and $Al_2O_3$ remain unchanged), thus lowering the coercive voltage and increasing the remnant polarization $P_r$. Additionally, more TiAl may lead to a lower electrode resistivity. The resistivities measured on a 4-point probe are 1.05E-1, 1.55E-5, 8.4E-6 $\Omega$.m for 2.5, 5.0, and 7.5 nm TiAl insertion layer thickness, respectively.

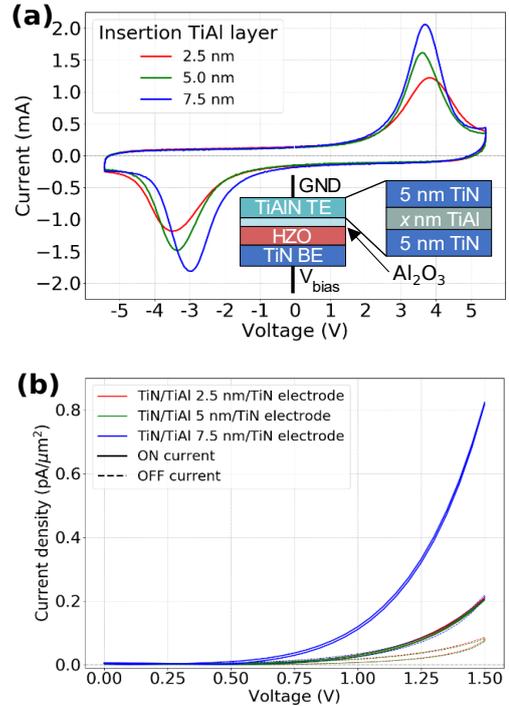

Fig. 3. (a) I-V switching characteristics measured at +/- 5.5 V, 1 kHz (waking up after 1e4 cycles at 5 V, 10 kHz) and (b) DC-IV on-current characteristic of the Al-doped TE samples.

The double-sweep I-V curves of three devices are shown in Fig. 3 b for the forward and backward sweep direction. With increasing TiAl thickness, the ON current rises due to the smaller work function, resulting in smaller potential barriers and a larger tunneling current. Obviously, the OFF current also increases. However, since in the differential synaptic circuit the decisive difference voltage $\Delta V_1 - \Delta V_2$ of the nodes $n_1$ and $n_2$ depends on the difference $I_{ON} - I_{OFF}$ rather than the current ratio $I_{ON} / I_{OFF}$, this effect is not detrimental. The $I_{ON} - I_{OFF}$ difference increased from <0.1 pA/µm² for the standard sample (see Fig. 2) to 0.6 pA/µm² for the TE with a 7.5 nm TiAl interlayer.

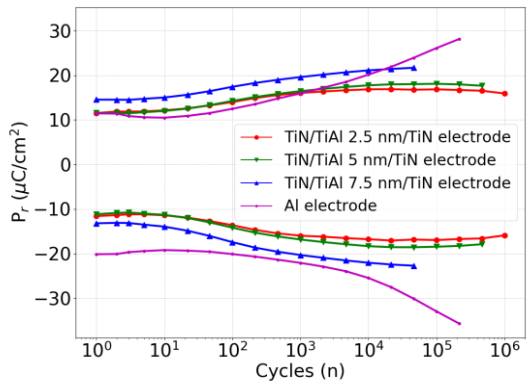

Fig. 4. Endurance cycling of the FTJ devices with Al-TE and TiN/TiAl/TiN TEs, cycled at +/- 5.5 V, 100 kHz; measured at 10 kHz, +/- 5.5 V.

Finally, the endurance cycling measurements of the Al-doped TE FTJ samples (see Fig. 4) revealed a much better reliability of the device compared to the Al-top electrode. In the case of Al-only, a high leakage after switching at cycles above $10^4$ obscures the polarization switching and could lead to unstable device behavior, whereas the $P_r$ for devices with TiAlN electrodes is seen to stabilize.

## IV. CONCLUSION

In this paper we have explored the possibility of using multiple FTJs in a differential synaptic cell for massive parallel neuromorphic computing. By analyzing the circuit considerations, we showed that due to the simultaneous scaling of both current density and self-capacitance, the voltage difference $\Delta V$ between the On- and Off-states is not improved by adding additional devices. From this, we conclude that the most important figure of merit for characterizing such multi-FTJ cells is the difference in current $\Delta I = I_{ON} - I_{OFF}$, with $\Delta V \propto \Delta J_{FTJ}$.

To achieve a large $\Delta I$, work function engineering of the TE can be applied. We have investigated Al, with a smaller work function than the standard TiN electrodes. However, pure Al as a TE leads to problems in reliability. We have shown that by instead doping Al into the stable TiN electrodes, $\Delta I$ can be increased from <0.1 pA/µm² for the standard sample to 0.6 pA/µm².

ACKNOWLEDGMENT

This work was supported by European Union through the BeFerroSynaptic project, Grant 871737 and by the DFG through the Memristec priority program, in the scope of the project ReLoFemRis (SL 305/2-1).